\documentclass[aps, prl, onecolumn]{revtex4}     
\usepackage{amsmath} \usepackage{amsfonts} \usepackage{amssymb}
\usepackage{wasysym}
\usepackage[dvipsnames]{xcolor}
\usepackage{graphicx}
\usepackage{multirow}
\usepackage{here}
\usepackage{epsfig}
\usepackage{epstopdf}
\usepackage{soul}
\usepackage{hyperref}
\usepackage{float}
\usepackage{mdframed}

\newcommand{\keV}{\mathinner{\mathrm{keV}}}
\newcommand{\MeV}{\mathinner{\mathrm{MeV}}}
\newcommand{\GeV}{\mathinner{\mathrm{GeV}}}
\newcommand{\TeV}{\mathinner{\mathrm{TeV}}}
\newcommand{\mpl}{M_\mathrm{Pl}}
\newcommand{\invMpc}{\mathinner{\mathrm{Mpc}^{-1}}}

\newcommand{\fref}[1]{Figure~\ref{#1}}

\renewcommand{\d}[1]{\mathinner{{\rm d}#1}}
\newcommand{\fn}[2]{\mathinner{#1\mathopen{\left(#2\right)}}}

\newcommand{\eq}[1]{Eq.~(\ref{#1})}

\definecolor{cadmiumgreen}{rgb}{0.0, 0.42, 0.24}

\begin{document}

\title{Observable Small-scale Effects of Thermal Inflation\footnote{To appear in the Proceedings of NDM-2020, Hurghada, Egypt. }}

\author{Heeseung Zoe}
\affiliation{School of Undergraduate Studies,
College of Transdisciplinary Studies, DGIST,
Daegu 42988, Republic of Korea}

\begin{abstract}
Thermal inflation, a brief low energy inflation after the primordial inflation, resolves the moduli problem in the context of supersymmetric cosmology. In the thermal inflation scenario, the primordial power spectrum is modestly redshifted on large scales, but suppressed by a factor of 1/50 on scales smaller than the horizon size at the beginning of thermal inflation. We compare the thermal inflation model with the warm dark matter and $\Lambda$CDM scenarios by studying CMB spectral distortions, halo abundances, and 21cm hydrogen lines.
\end{abstract}

\maketitle


\section{Moduli Problem}
\setcounter{equation}{0}
\renewcommand{\theequation}{1.\arabic{equation}}

Moduli fields are generic in supersymmetric/string theories. They couple to matter with only gravitational strengths, and the decay rate is estimated by $\Gamma_{\Phi} \sim m_\Phi^3/\mpl^2$, where $m_\Phi$ is the moduli mass and $\mpl = 2.44 \times 10^{18}\GeV$ is the reduce Planck mass. The lifetime of moduli is estimated by 
\begin{equation}
\label{lifetime}
\tau_\Phi \sim \tau_{\rm uni}\left( \frac{100 \MeV}{m_\Phi} \right)^2
\end{equation}
where $\tau_{\rm uni} \sim 4.3 \times 10^{17} {\rm sec} $ is the age of universe. 

When the energy scale of the universe is larger than the supersymmetry breaking scale, the moduli potential is the form of $V_\Phi \sim H^2 \left( \Phi - \Phi_1 \right)^2$. As the universe expands, the energy scale of the universe decreases, and the moduli potential turns to its vacuum form of $V_\Phi \sim m_\Phi^2 \left( \Phi - \Phi_2 \right)^2$.
For $ \Phi_0 \equiv \left| \Phi_1 - \Phi_2 \right| \sim \mpl$, the moduli would be oscillating with the Planckian amplitude. If the moduli decay out right after its oscillation, the entropy release is estimated to be
\begin{equation}
\label{entropy}
\frac{n_\Phi}{s} \sim \frac{\Phi_0^2}{ 10 \mpl^{3/2} m_\Phi^{1/2}}~.
\end{equation}
This value remains constant unless there is no dramatic event with the moduli, and can be very large depending on the moduli mass. 

The moduli problem is two-folded: (1) For $\tau_\Phi < \tau_{\rm uni}$, the moduli decay might release a huge amount of entropy in the universe and dilute its contents. Hence, it could disturb the Big Bang Nucleosynthesis (BBN). (2) For $\tau_\Phi > \tau_{\rm uni}$, the moduli might still be oscillating around the minimum, and the energy stored in the oscillations would overclose the universe. 

\begin{figure}[ht]
\centering
\includegraphics[height=2.8in]{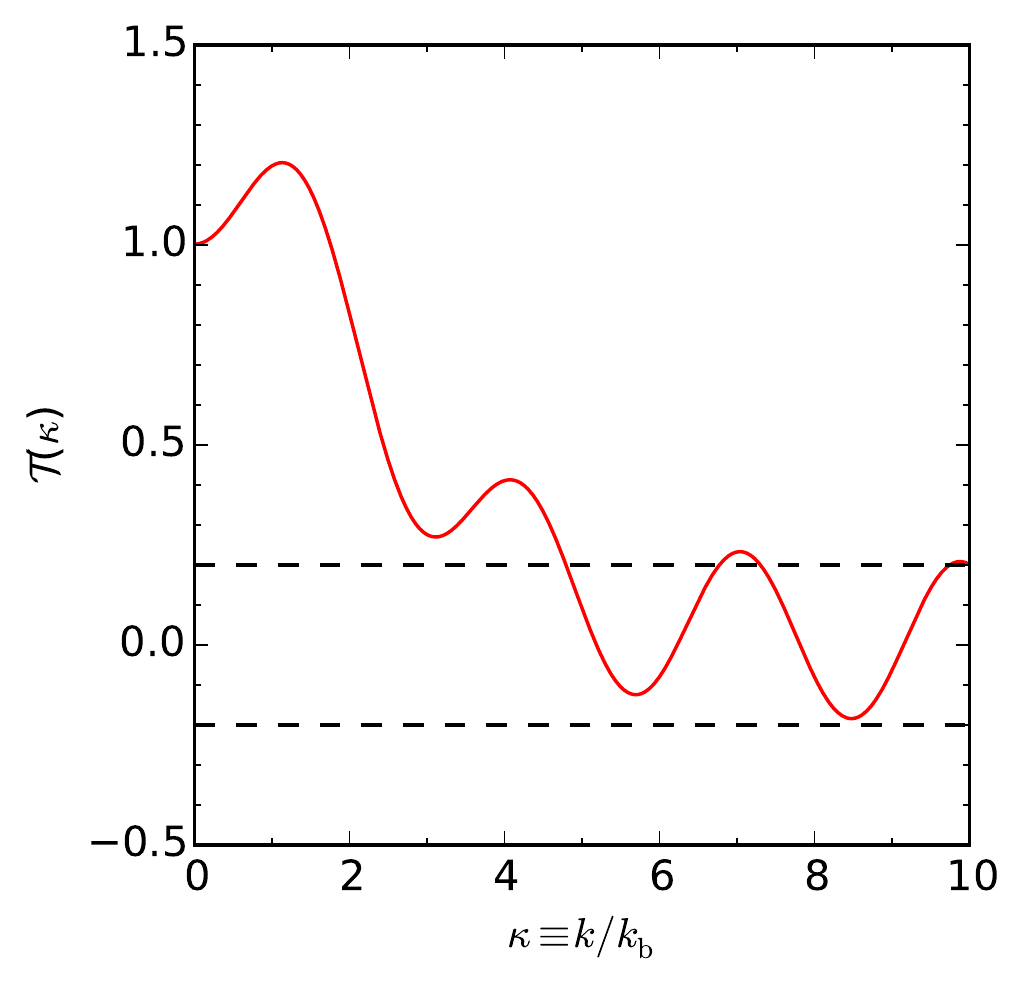}
\includegraphics[height=2.75in]{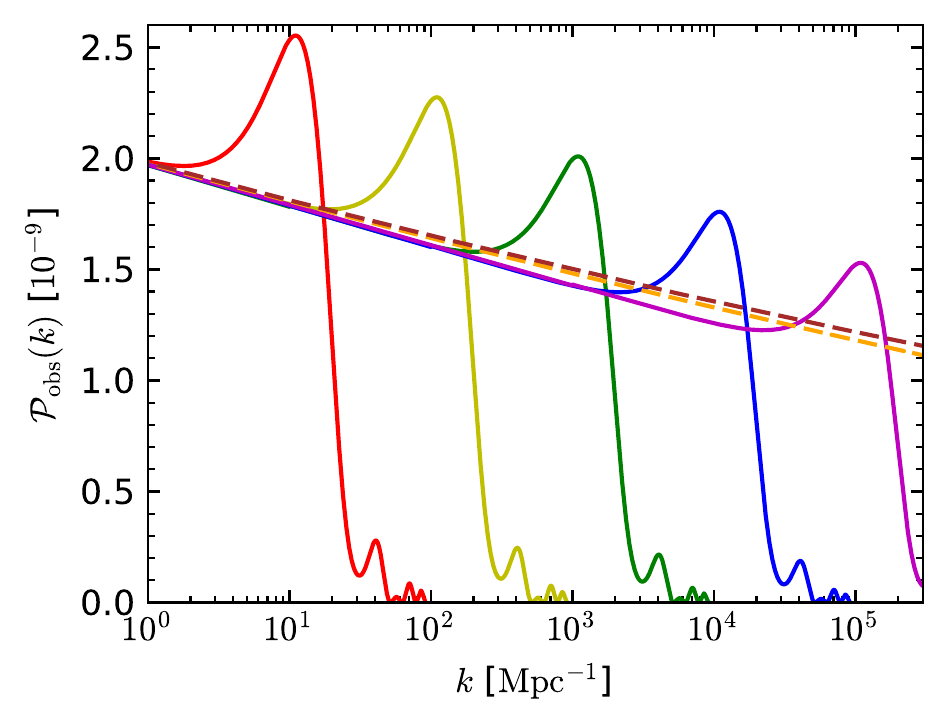}
\caption{Left: The transfer function of thermal inflation of \eq{transfer} (Credit: \cite{Hong:2015oqa}). Right: The power spectrum of thermal inflation with $k_{\rm b}/\invMpc = {\color{red}10}$, $\color{Goldenrod}10^2$, $\color{cadmiumgreen}10^3$, $\color{blue}10^4$, $\color{violet}10^5$. The dashed lines are the power spectra without thermal inflation. $\mathcal{N}_{\rm unkown} = {\color{orange}0}$, $\color{brown}-13$, $- \infty$ represent the amount of inflation between the beginning of the moduli domination and the end of the primordial inflation (Credit: \cite{Cho:2017zkj}).  }
\label{PS}
\end{figure}

\section{Thermal Inflation}
\setcounter{equation}{0}
\renewcommand{\theequation}{2.\arabic{equation}}

The moduli problem is discussed in the context of the gravity-mediated suersymmetry breaking model in the thermal inflation scenario \cite{Lyth:1995hj,Lyth:1995ka}. 
For $m_\phi \sim 1 \TeV $, we estimate
\begin{equation}
\tau_\Phi \sim 10^{-12} \times \tau_{\rm uni}
\end{equation}
from \eq{lifetime} and 
\begin{equation}
\label{MP}
\frac{n_\Phi}{s} \sim 10^7
\end{equation}
from \eq{entropy}. The entropy release is so huge that it conflicts with the observational bounds for successful BBN, $10^{-14}$ to $10^{-12}$ \cite{Kawasaki:2004qu}. 

In the thermal inflation scenario, the moduli problem is resolved by introducing a flaton field, a scalar field generated by a supersymmetric flat direction with considering thermal effects on it. The flaton potential becomes $V_\phi = V_0 +\frac{1}{2} \left( g^2 T^2 - m_\phi^2 \right)\phi^2 + \cdots $ around $\phi = 0$, where $g$ is the thermal coupling, and $T$ is the temperature of the universe. When the temperature is very high, the flaton is held at the origin. As the temperature decreases with the expansion of the universe, the vacuum energy $V_0$ drives the inflationary phase, and the pre-existing moduli can be diluted by $\Delta_{\rm TI} \sim 10^{-12}$. In the scenario having successful Affleck-Dine baryogenesis, there can be found the additional dilution factor $\Delta_{\rm AD} \sim 10^{-8}$. Hence, we can reduce the entropy release of \eq{MP} sufficiently to the observational bounds. However, the moduli can be regenerated due to the misalignment of the potential at the end of thermal inflation by
\begin{equation}
\frac{n_\Phi}{s} \sim \frac{V_0 T_{\rm reh}}{m_\Phi^2 \mpl^2} \sim 10^{-14} \left( \frac{V_0^{1/4}}{10^8 \GeV} \right)^4\left( \frac{T_{\rm reh}}{\GeV}  \right)
\end{equation}
where $T_{\rm reh}$ is the reheating temperature of the flaton. If the energy scale of thermal inflation is low enough, $V_0^{1/4} \lesssim 10^8 \GeV $, it would not conflict with the observational bounds for BBN.

\section{Power spectrum of thermal inflation}

\setcounter{equation}{0}
\renewcommand{\theequation}{3.\arabic{equation}}

The outline of the thermal inflation scenario consists of the following five steps:

\begin{center}
\noindent\fbox{%
    \parbox{0.95\textwidth}{%

\begin{enumerate}
\item Primordial inflation produces the (nearly) scale-invariant power spectrum. 

\item The moduli $\Phi$ dominates over the universe.

\item Thermal inflation gets started to resolve the moduli problem by the flaton $\phi$ with thermal effects on it. 

\item The flaton oscillates around the its minimum and dominates over the universe.

\item As the inflaton decays out, it reheat the universe, and the radiation dominates over the universe.
\end{enumerate}
    }%
}

\end{center}

In \cite{Hong:2015oqa}, we see that cosmological perturbations on large scales, corresponding to cosmic microwave background (CMB) or large scale structure (LSS) observations, remain preserving while perturbations on small scales are suppressed.
At the step 1, the primordial power spectrum is given by 
$\frac{\d \ln P_{\rm pri}}{\d \ln k} = - \frac{c}{\mathcal{N}}$, 
where $c$ is a constant and $\mathcal{N} \equiv \ln \frac{k}{a_{\rm end}H_{\rm end}}$ is the amount of inflation from the `end' of the primordial inflation.  
After the primordial inflation, there is an unknown era which is approximately described by the equation of state of a single component, $p = w\rho$ with $0 \leqslant \omega \leqslant 1/3$.
Now we summarize the change of the primordial power spectrum from the step 2 to the step 4 of the above by a transfer function 
\begin{equation}
\label{transfer}
\fn{\mathcal{T}}{\frac{k}{k_{\rm b}}}=\cos \left[\frac{k}{k_{\rm b}} \int_0^\infty \frac{d\alpha}{\sqrt{\alpha(2+\alpha^3)}} \right] 
+ 6 \frac{k}{k_{\rm b}} \int_0^\infty \frac{d\gamma}{\gamma^3} \int_0^\gamma d\beta \left( \frac{\beta}{2+\beta^3}\right)^\frac{3}{2}
\sin \left[ \frac{k}{k_{\rm b}} \int_\gamma^\infty \frac{d\alpha}{\sqrt{\alpha(2+\alpha^3)}} \right]
\end{equation}
where the characteristic scale $k_{\rm b}$ is estimated by 
\begin{equation}
\label{kb}
k_{\rm b} \simeq 3 \times 10^3\, \invMpc \left( \frac{e^{20}}{e^{N_{\rm TI}}} \right)
\left( \frac{\fn{g_*}{T_{\rm reh}}}{10^2}  \right)^{1/{12}}
\left( \frac{T_{\rm reh}}{\GeV}  \right)^{1/3}
\left( \frac{V_0^{1/4}}{10^7 \GeV}  \right)^{2/3}
  \, . 
\end{equation}
with the efolds during thermal inflation $N_{\rm TI}$.  
In \eq{kb}, $N_{\rm TI} \sim 10$ to $15$ is enough to resolve the moduli problem. However, there is no theoretical upper bound for $N_{\rm TI}$, and if we include multiple thermal inflations, $N_{\rm TI}$ should get larger. 
Hence, $N_{\rm TI}$ can be larger as long as it is not consistent with the large scale observations such as CMB. 

The resulting power spectrum is given by 
\begin{equation}
\label{power}
\fn{P}{k} = \fn{\mathcal{T}^2}{k/k_\mathrm{b}} \times \fn{P_\mathrm{pri}}{k}~.
\end{equation}
The transfer function of \eq{transfer} is shown in the left panel of \fref{PS} and the resulting power spectra of \eq{power} are shown in the right of \fref{PS}.
The crucial feature of the thermal inflation is that the power spectrum is suppressed by $1/50$ at $k \gtrsim k_{\rm b}$. Hence, thermal inflation can be explored by cosmological or astrophysical observations on such small scales.

\section{Observational Effects on Small Scales}

\begin{figure}[ht]
\centering
\includegraphics[height=2.8in]{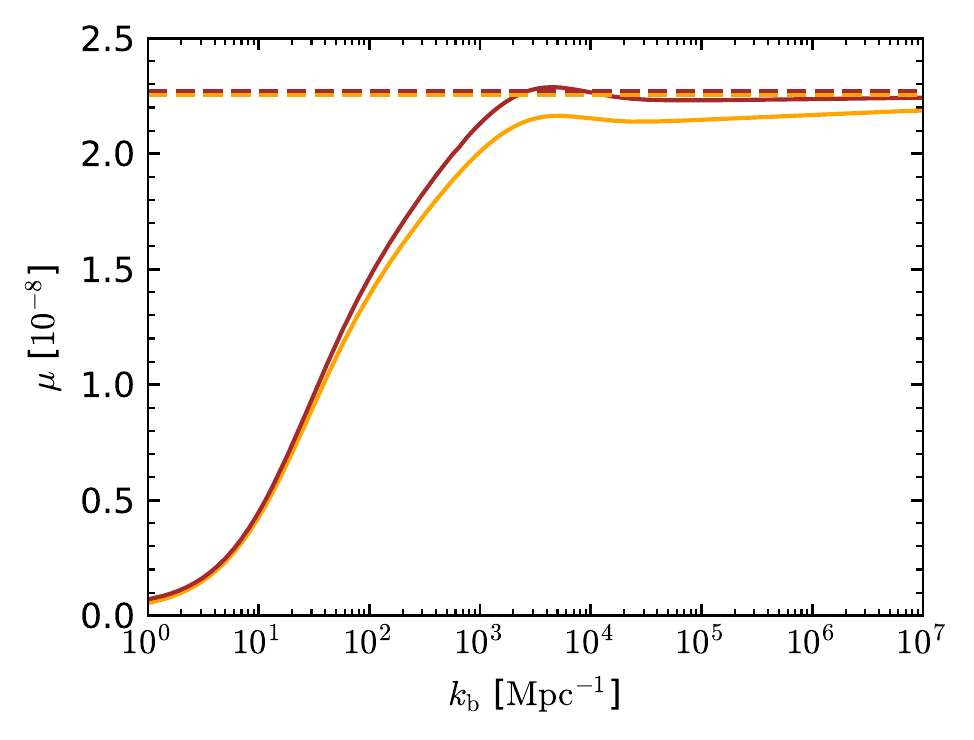}
\includegraphics[height=2.8in]{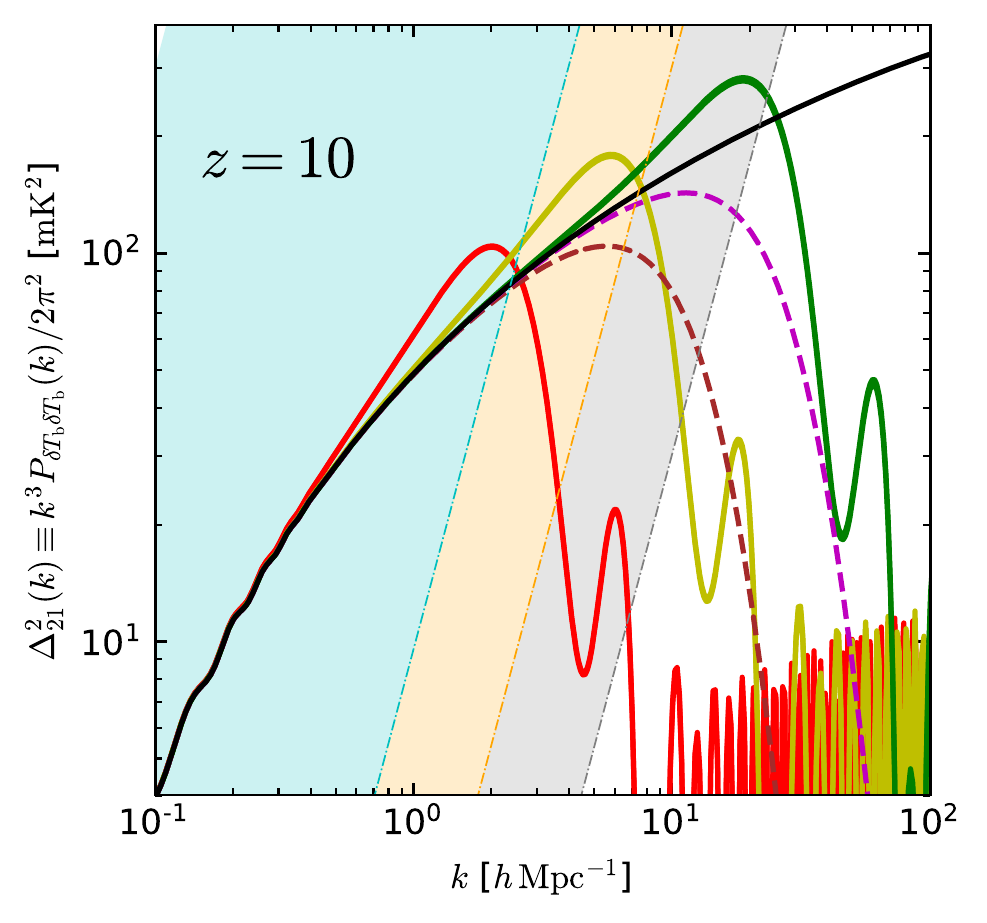}
\caption{Left: CMB spectral $\mu$-distortion with $\mathcal{N}_{\rm unkown} = {\color{orange}0}$, $\color{brown}-13$, $- \infty$ (Credit: \cite{Cho:2017zkj}). Right: The 21-cm power spectra of thermal inflation ($k_{\rm b} =$ {\color{red} $1\invMpc$ (red)}, {\color{Goldenrod} $3\invMpc$ (yellow)},  {\color{cadmiumgreen} $5\invMpc$  (green)}), warm dark matter ( $m_{\rm FD} =$ {\color{brown} $1 \keV$ (brown)}, {\color{magenta} $2 \keV$ (magenta)}), and $\Lambda$CDM (black) scenarios just before the epoch of reionization. The shaded regions imply the power spectra above the thermal noise from the modified SKA configuration with $100~{\rm deg^2}$ sky coverage. Exposure times are ${\color{cyan}10^3}$, ${\color{orange}10^4}$, and ${\color{gray}10^5}$-hours in SKA1-LOW, and ${\color{cyan}10^2}$, ${\color{orange}10^3}$, and ${\color{gray}10^4}$-hours in SKA2-LOW \citep{Greig:2015zra} (Credit: \cite{Hong:2017knn}).
}
\label{obs}
\end{figure}

The power spectrum of thermal inflation can be explored by the following three methods:

\begin{itemize}

\item  CMB spectral distortion is generated by the dissipation of primordial density fluctuations due to the Silk damping of acoustic waves \cite{Chluba:2012we}. It is sensitive to the power spectrum on the scales $1 \invMpc \lesssim  k \lesssim 10^4 \invMpc$ and can be used to probe the thermal inflation power spectrum. In the left panel of \fref{obs}, the spectral $\mu$-distortion of thermal inflation falls well below the value of the standard $\Lambda$CDM scenario at $k_{\rm b} \lesssim 10^3 \invMpc$ \cite{Cho:2017zkj}. 

\item  The suppression of the thermal inflation power spectrum gives a possible way of explaining the missing satellite problem \cite{Kauffmann:1993gv}. The substructure of the galaxies can be useful in studying the thermal inflation cosmology compared to the warm dark matter scenario having the similar effects on small scales \cite{Hong:2017knn}. In \cite{Leo:2018kxp}, the halo mass function of N-body simulation of the thermal inflation scenario is distinguished from that of the warm dark matter and $\Lambda$CDM scenarios.

\item The hydrogen 21-cm line background can be used to test the matter power spectrum of the thermal inflation scenario. In the right panel of \fref{obs}, the  Square Kilometre Array (SKA) would be able to probe the differences of the hydrogen distributions between the thermal inflation, warm dark matter, and $\Lambda$CDM scenarios \cite{Hong:2017knn}.

\end{itemize}

\section*{Acknowledgements}
The author thanks Sungwook Hong and Ewan Stewart for helpful discussions. 
This work is supported by the DGIST  UGRP grant.

\bibliographystyle{unsrt}

\end{document}